\documentstyle[12pt]{article}
\def\be {\begin{equation}}
\def\ee {\end{equation}}
\def\ba {\begin{eqnarray}}
\def\ea {\end{eqnarray}}

\def\bi {\begin{itemize}}
\def\ei {\end{itemize}}
\begin{document}
\def\bea{\begin{eqnarray}}
\def\eea{\end{eqnarray}}
\title{\bf {Bulk-Brane Interaction and Holographic Dark Energy}}
 \author{M.R. Setare  \footnote{E-mail: rezakord@ipm.ir}
  \\{Department of Science,  Payame Noor University. Bijar. Iran}}
\date{\small{}}

\maketitle
\begin{abstract}
In this paper we consider the bulk-brane interaction to obtain
the equation of state  for the holographic energy density
 in non-flat universe enclosed by
 the event horizon measured from the
 sphere of horizon named $L$. We
assumes that the cold dark matter energy density on the brane is
conserved, but the holographic dark  energy density on the brane is
not conserved due to brane-bulk energy exchange. Our calculation
show, taking $\Omega_{\Lambda}=0.73$ for the present time, the lower
bound of $w_{\rm \Lambda}^{eff}$ is $-0.9$. This implies that one
can not generate phantom-like equation of state from an interacting
holographic dark energy model in non-flat universe.
 \end{abstract}

\newpage

\section{Introduction}

One of the most important problems of cosmology, is the problem of
so-called dark energy (DE). The type Ia supernova observations
suggests that the universe is dominated by dark energy with
negative pressure which provides the dynamical mechanism of the
accelerating expansion of the universe \cite{{per},{gar},{ries}}.
The strength of this acceleration is presently matter of debate,
mainly because it depends on the theoretical model implied when
interpreting the data. Most of these models are based on dynamics
of a scalar or multi-scalar fields (e.g quintessence
\cite{{rat},{zlat}} and quintom model of dark energy,
respectively). \\
An approach to the problem of DE arises from holographic Principle
that states that the number of degrees of freedom related directly
to entropy scales with the enclosing area of the system. It was
shown by 'tHooft and Susskind \cite{hologram} that effective local
quantum field theories greatly overcount degrees of freedom
because the entropy scales extensively for an effective quantum
field theory in a box of size $L$ with UV cut-off $ \Lambda$. As
pointed out by \cite{myung}, attempting to solve this problem,
Cohen {\it et al.} showed \cite{cohen} that in quantum field
theory, short distance cut-off $\Lambda$ is related to long
distance cut-off $L$ due to the limit set by forming a black hole.
In other words the total energy of the system with size $L$ should
not exceed the mass of the same size black hole i.e. $L^3
\rho_{\Lambda}\leq LM_p^2$ where $\rho_{\Lambda}$ is the quantum
zero-point energy density caused by UV cutoff $\Lambda$ and $M_P$
denotes Planck mass ( $M_p^2=1/{8\pi G})$. The largest $L$ is
required to saturate this inequality. Then its holographic energy
density is given by $\rho_{\Lambda}= 3c^2M_p^2/8\pi L^2$ in which
$c$ is free dimensionless parameter and coefficient 3 is for
convenience.

 As an application of Holographic principle in cosmology,
 it was studied by \cite{KSM} that consequence of excluding those degrees of freedom of the system
 which will never be observed by that effective field
 theory gives rise to IR cut-off $L$ at the
 future event horizon. Thus in a universe dominated by DE, the
 future event horizon will tend to constant of the order $H^{-1}_0$, i.e. the present
 Hubble radius. The consequences of such a cut-off could be
 visible at the largest observable scales and particulary in the
 low CMB multipoles where we deal with discrete wave numbers. Considering the power spectrum in finite
 universe as a consequence of holographic constraint, with different boundary
 conditions, and fitting it with LSS, CMB and supernova data, a cosmic duality between dark energy equation of state
 and power spectrum is obtained that can describe the low $l$ features extremely
 well.

 Based on cosmological state of holographic principle, proposed by Fischler and
Susskind \cite{fischler}, the Holographic Model of Dark Energy
(HDE) has been proposed and studied widely in the
 literature \cite{miao,HDE}. In \cite{HG} using the type Ia
 supernova data, the model of HDE is constrained once
 when c is unity and another time when c is taken as free
 parameter. It is concluded that the HDE is consistent with recent observations, but future observations are needed to
 constrain this model more precisely. In another paper \cite{HL},
 the anthropic principle for HDE is discussed. It is found that,
 provided that the amplitude of fluctuation are variable the
 anthropic consideration favors the HDE over the cosmological
 constant.

 In HDE, in order to determine the proper and well-behaved system's IR cut-off, there are some
difficulties that must be studied carefully to get results adapted
with experiments that claim our universe has accelerated
expansion. For instance, in the model proposed by \cite{miao}, it
is discussed that considering particle horizon, as the IR
cut-off, the HDE density reads to be
 \be
  \rho_{\Lambda}\propto a^{-2(1+\frac{1}{c})},
\ee
 that implies $w>-1/3$ which does not lead to accelerated
universe. Also it is shown in \cite{easther} that for the case of
closed
universe, it violates the holographic bound.\\
The problem of taking apparent horizon (Hubble horizon) - the
outermost surface defined by the null rays which instantaneously
are not expanding, $R_A=1/H$ - as the IR cut-off in the flat
universe, was discussed by Hsu \cite{Hsu}. According to Hsu's
argument, employing Friedman equation $\rho=3M^2_PH^2$ where
$\rho$ is the total energy density and taking $L=H^{-1}$ we will
find $\rho_m=3(1-c^2)M^2_PH^2$. Thus either $\rho_m$ and
$\rho_{\Lambda}$ behave as $H^2$. So the DE results pressureless,
since $\rho_{\Lambda}$ scales as like as matter energy density
$\rho_m$ with the scale factor $a$ as $a^{-3}$. Also, taking
apparent horizon as the IR cut-off may result the constant
parameter of state $w$, which is in contradiction with recent
observations implying variable $w$ \cite{varw}.
 On the other hand taking the event horizon, as
 the IR cut-off, gives the results compatible with observations for flat
 universe.

 It is fair to claim that simplicity and reasonability of HDE provides
 more reliable frame to investigate the problem of DE rather than other models
proposed in the literature\cite{cosmo,quint,phant}. For instance
the coincidence or "why now" problem is easily solved in some
models of HDE based on this fundamental assumption that matter and
holographic dark energy do not conserve separately, but the matter
energy density
decays into the holographic energy density \cite{interac}.\\
Some experimental data has implied that our universe is not a
perfectly flat universe and recent papers have favored the
universe with spatial curvature \cite{curve}. As a matter of
fact, we want to remark that although it is believed that our
universe is flat, a contribution to the Friedmann equation from
spatial curvature is still possible if the number of e-foldings
is not very large \cite{miao2}. Defining the appropriate
distance, for the case of non-flat universe has another story.
Some aspects of the problem has been discussed in
\cite{miao2,guberina}. In this case, the event horizon can not be
considered as the system's IR cut-off, because for instance, when
the dark energy is dominated and $c=1$, where $c$ is a positive
constant, $\Omega_\Lambda=1+ \Omega_k$, we find $\dot R_h<0$,
while we know that in this situation we must be in de Sitter
space with constant EoS. To solve this problem, another distance
is considered- radial size of the event horizon measured on the
sphere of the horizon, denoted by $L$- and the evolution of
holographic model of dark energy in non-flat universe is
investigated.
\\
Motivated by string/M theory, the AdS/CFT correspondence, and the
hierarchy problem of particle physics, braneworld models were
studied actively in recent years \cite{Hora96}-\cite{Rand99}. In
this models, our universe is realized as a boundary of a higher
dimensional spacetime.\\
In this paper we turn our attention to the interaction between the
bulk and the brane, which is  non-trivial aspect of brane world
theories. In particular, we will discuss the flow of energy onto or
away from the brane--universe. Then,  using the holographic model of
dark energy in non-flat universe, we obtain equation of state for
interacting holographic dark energy density in a universe enveloped
by  $L$ as the system's IR cut-off. At first we review the formalism
of bulk-brane energy exchange, then we applies this material to a
brane-world cosmology containing cold dark matter (CDM), DE, and
spatial curvature. The DE is assumed to be of HDE variety. We
assumes that the CDM energy density on the brane is conserved, but
the the holographic dark  energy density on the brane is not
conserved due to brane-bulk energy exchange.
\section{Bulk-Brane Energy Exchange }
We consider the following gravitational brane-bulk action
\begin{equation}\label{eq1}
    S=\int d^5 x
    \sqrt{-G}\left(\frac{R_5}{2\kappa_5^2}-\Lambda_5+L_{B}^{m} \right)+ \int
    d^4 x \sqrt{-g}\left(-\sigma + L_{b}^{m} \right),
\end{equation}
where $R_5$ is the curvature scalar of the five-dimensional
metric, $\Lambda_5$ is the bulk cosmological constant and $\sigma$
is the brane tension, $L_{B}^{m}$ and $L_{b}^{m}$ are the matter
Lagrangian in the bulk and on the brane respectively. We consider
an ansatz for the metric of the form
\begin{equation}\label{eq2}
    ds^2 = -n^2(t,y)\,dt^2 + a^2(t,y)\,\gamma_{ij}\,dx^i dy^j +
    b^2(t,y)\,dy^2.
\end{equation}
where $\gamma_{ij}$ is the metric for the  maximally symmetric
three-dimensional space. The non-zero components of Einstein
tensor can be written as \cite{{21},{kir}} \be G_{00}=3 \left[
\frac{ \dot a}{a} \left( \frac{ \dot a}{a} + \frac{ \dot b}b
\right) - \frac{n^2}{b^2} \left( \frac{a^{ \prime \prime }}a +
\frac{a^{ \prime }}a \left( \frac{a^{ \prime}}a-
\frac{b^{\prime}}b \right) \right) + k \frac{n^2}{b^2}
\right]\label{g00} \ee
 \bea G_{ij}&=&
\frac{a^2}{b^2}\gamma_{ij} \left[\frac{a^{ \prime}}a \left(
\frac{a^{ \prime}}a+2\frac{n^{
\prime}}n\right)-\frac{b^{\prime}}b\left(
\frac{n^{\prime}}n+2\frac{a^{ \prime}}a\right)+2\frac{a^{\prime
\prime}}a+ \frac{n^{\prime \prime}}n\right] \nonumber \\
& +&\frac{a^2}{n^2}\gamma_{ij}\left[ \frac{\dot
a}a\left(-\frac{\dot a}a+2\frac{\dot n}n\right)-2\frac{\ddot a}a
+\frac{\dot b}b\left(-2\frac{\dot a}a+\frac{\dot n}n
\right)-\frac{\ddot b}b \right] -k\gamma_{ij}, \label{gij} \eea
\be G_{05}=3\left( \frac{n^{\prime}}n\frac{\dot a}a
+\frac{a^{\prime}}a\frac {\dot b}b - \frac{\dot a^{\prime}}a
\right), \label{g05}
 \ee
 \be
G_{55}=3\left[\frac{a^{\prime}}a\left(\frac{a^{\prime}}a+
\frac{n^{\prime}}n \right)-\frac{b^2}{n^2}\left(\frac{\dot
a}a\left(\frac{ \dot a}a-\frac{\dot n}n \right)+\frac{\ddot
a}a\right)-k\frac{b^2}{a^2}\right], \label{g55}
 \ee

 where $k$ denotes the curvature of space k=0,1,-1 for flat, closed
and open universe respectively. The three dimensional brane is
assumed at $y=0$. The Einstein equations are $G_{\mu \nu}=
\kappa^2_5 T_{\mu\nu}$, where
 the stress-energy momentum tensor has  bulk and brane
components and can be written as
\begin{equation}
\label{eq7}
    T^{\mu}_{\phantom{\mu}\nu} = T^{\mu}_{\phantom{\mu}\nu}|_{\sigma,\,b}
    + T^{\mu}_{\phantom{\mu}\nu}|_{m,b} +
    T^{\mu}_{\phantom{\mu}\nu}|_{\Lambda,\,B} +
    T^{\mu}_{\phantom{\mu}\nu}|_{m,B},
\end{equation}
where
\begin{eqnarray}
  T^{\mu}_{\phantom{\mu}\nu}|_{\sigma,\,b} &=& \frac{\delta(y)}{b}\,
  \mathrm{diag}(-\sigma,\,-\sigma,\,-\sigma,\,-\sigma,\,0), \label{eq8}\\
  T^{\mu}_{\phantom{\mu}\nu}|_{\Lambda,\,B} &=& \mathrm{diag}(-\Lambda_5,
  \,-\Lambda_5,\,-\Lambda_5,\,-\Lambda_5,\,-\Lambda_5), \label{eq9}\\
  T^{\mu}_{\phantom{\mu}\nu}|_{m,\,b} &=&
  \frac{\delta(y)}{b}\,\mathrm{diag}(-\rho,\,p,\,p,\,p,\,0),\label{eq10}
\end{eqnarray}
$\rho$ and $p$ are energy density and pressure on the brane,
respectively. Integrating eqs.(\ref{g00}, \ref{gij}) with respect
to $y$ around $y=0$ give the following jump conditions
\begin{eqnarray}
  a'_{+} &=& -\,a'_{-} = -\frac{\kappa_5^2}{6}\,a_0\,b_0\,(\sigma + \rho), \label{eq11}\\
  n'_{+} &=& -\,n'_{-} = \frac{\kappa_5^2}{6} \,b_0\,n_0\,(-\sigma + 2\rho
  +3p),
   \label{eq12}
 \end{eqnarray}
 Employing
Eqs.~(\ref{eq11}) and (\ref{eq12}), we can derive\cite{kir}
\begin{eqnarray}
  \dot{\rho} + 3 \frac{\dot{a}_0}{a_0}(\rho+p) &=& -\frac{2
n^2_0}{b_0}T^0_{\phantom{0}5},
  \label{eq15} \\
  \frac{1}{n_0^2} \left[ \frac{\ddot{a}_0}{a_0} + \left( \frac{\dot{a}_0}{a_0} \right)^2
   -\frac{\dot{a}_0\,\dot{n}_0}{a_0\,n_0} \right] + \frac{k}{a_0^2} &=&
  \frac{\kappa_5^2}{3} \left( \Lambda_5 + \frac{\kappa_5^2\,\sigma^2}{6}\right)
  \nonumber\\
    & & - \frac{\kappa_5^4}{36} \left[ \sigma(3p-\rho)+\rho(3p+\rho)
  \right] - \frac{\kappa_5^2}{3}\,T^5_{\phantom{5}5}.
  \label{eq16}
\end{eqnarray}
where $T_{05}$ and $T_{55}$ are the 05 and 55 components of
$T_{\mu\nu}|_{m,\,B}$ evaluated on the brane. In order to derive
a solution that is largely independent of the bulk dynamics, we
can neglect $T^5_{\phantom{5}5}$ term by assuming that the bulk
matter relative to the bulk vacuum energy is much less than the
ratio of the brane matter to the brane vacuum energy~\cite{19}.
Considering this approximation and concentrating on the
low-energy region with $\rho / \sigma \ll 1$,
Eqs.~(\ref{eq15},\ref{eq16}) can be simplified into \cite{cai}
\begin{eqnarray}
  \dot{\rho} + 3 H(1+w)\rho &=& -2\,T^0_{\phantom{0}5} = T \label{eq20} \\
  H^2 &=& \frac{8\pi G_4}{3} (\rho + \chi) - \frac{k}{a^2} + \lambda \label{eq21}\\
  \dot{\chi} + 4H\chi & \approx & 2
  \,T^0_{\phantom{0}5} =-T . \label{eq22}
\end{eqnarray}
Thus with the energy exchange $T$ between the bulk and brane, the
usual energy conservation is broken down. In the following we will
consider that there are two dark components in the universe, dark
matter and dark energy, $\rho = \rho_m + \rho_{\Lambda}$. Here we
assume that the adiabatic equation for the dark matter is
satisfied while it is violated for the dark energy due to the
energy exchange between the brane and the bulk,
\begin{eqnarray}
  \dot{\rho}_m + 3 H \rho_m &=& 0, \label{eq23}\\
  \dot{\rho}_{\Lambda} + 3 H (1+w_{\Lambda}) \rho_{\Lambda} &=& T. \label{eq24}
\end{eqnarray}
The interaction between bulk and brane is given by the quantity
$T=\Gamma \rho_{\Lambda}$, where $\Gamma$ is the rate of
interaction. Taking a ratio of two energy densities as
$u=\chi/\rho_{\rm \Lambda}$, the above equations lead to
\begin{equation}\label{ratio}
\dot{u}=3Hu[w_\Lambda-\frac{1}{3}-\frac{1+u}{u}\frac{\Gamma}{3H}]
\end{equation}

If we define
\begin{eqnarray}\label{eff}
w_\Lambda ^{\rm eff}=w_\Lambda-{{\Gamma}\over {3H}},
\end{eqnarray}
Then, the continuity eq.(\ref{eq24}) can be written in following
standard form
\begin{equation}
\dot{\rho}_\Lambda + 3H(1+w_\Lambda^{\rm eff})\rho_\Lambda = 0.
\label{definew}
\end{equation}
\\
We study the case of a universe with zero effective cosmological
constant, but with spatial curvature. A closed universe with a
small positive curvature ($\Omega_k\sim 0.01$) is compatible with
observations \cite{curve}. Define as usual
\begin{equation} \label{2eq9} \Omega_{\rm
m}=\frac{\rho_{m}}{\rho_{cr}},\hspace{1cm}\Omega_{\rm
\Lambda}=\frac{\rho_{\Lambda}}{\rho_{cr}}, \hspace{1cm}
\Omega_{\chi}=\frac{\chi}{\rho_{cr}},
\hspace{1cm}\Omega_{k}=\frac{k}{a^2H^2}
\end{equation}
where $\rho_{cr}=\frac{3H^2}{8\pi G_4}$. Now we can rewrite the
first Friedmann equation  (\ref{eq21}) as
\begin{equation} \label{2eq10} \Omega_{\rm m}+\Omega_{\rm
\Lambda}+\Omega_{\chi}=1+\Omega_{k}.
\end{equation}
Using Eqs.(\ref{2eq9},\ref{2eq10}) we obtain following relation for
ratio of energy densities $r$ as
\begin{equation}\label{ratio}
u=\frac{1+\Omega_{k}-\Omega_{\Lambda}-\Omega_{\rm
m}}{\Omega_{\Lambda}}
\end{equation}
In non-flat universe, our choice for holographic dark energy
density is
 \be
  \rho_\Lambda=\frac{3c^2}{8\pi G_4L^{2}}.
 \ee
As it was mentioned, $c$ is a positive constant in holographic
model of dark energy($c\geq1$)and the coefficient 3 is for
convenient. $L$ is defined as the following form:
\begin{equation}\label{leq}
 L=ar(t),
\end{equation}
here, $a$, is scale factor and $r(t)$ can be obtained from the
following equation
\begin{equation}\label{rdef}
\int_0^{r(t)}\frac{dr}{\sqrt{1-kr^2}}=\int_t^\infty
\frac{dt}{a}=\frac{R_h}{a},
\end{equation}
where $R_h$ is event horizon. For closed universe we have (same
calculation is valid for open universe by transformation)
 \be \label{req}
 r(t)=\frac{1}{\sqrt{k}} sin y.
 \ee
where $y\equiv \sqrt{k}R_h/a$. Using definitions
$\Omega_{\Lambda}=\frac{\rho_{\Lambda}}{\rho_{cr}}$, we get

\begin{equation}\label{hl}
HL=\frac{c}{\sqrt{\Omega_{\Lambda}}}
\end{equation}
Now using Eqs.(\ref{leq}, \ref{rdef}, \ref{req}, \ref{hl}), we
obtain \be \label{ldot}
 \dot L= HL+ a \dot{r(t)}=\frac{c}{\sqrt{\Omega_\Lambda}}-cos y,
\end{equation}
By considering  the definition of holographic energy density
$\rho_{\rm \Lambda}$, and using Eqs.( \ref{hl}, \ref{ldot}) one
can find:
\begin{equation}\label{roeq}
\dot{\rho_{\Lambda}}=-2H(1-\frac{\sqrt{\Omega_\Lambda}}{c}\cos
y)\rho_{\Lambda}
\end{equation}
Substitute this relation into Eq.(\ref{eq24}) and using
definition $T=\Gamma \rho_{\Lambda}$, we obtain
\begin{equation}\label{stateq}
w_{\rm \Lambda}=-(\frac{1}{3}+\frac{2\sqrt{\Omega_{\rm
\Lambda}}}{3c}\cos y-\frac{\Gamma}{3H}).
\end{equation}
Here we choose the following ansatz \cite{wang}
\begin{equation}\label{decayeq}
\Gamma=3b^2(1+u)H
\end{equation}
with  the coupling constant $b^2$. Using Eq.(\ref{ratio}), the
above  equation take following form
\begin{equation}\label{decayeq2}
\Gamma=3b^2H\frac{(1+\Omega_{k}-\Omega_{\rm m})}{\Omega_{\Lambda}}
\end{equation}
Substitute this relation into Eq.(\ref{stateq}), one finds the
holographic energy equation of state
\begin{equation} \label{3eq4}
w_{\rm \Lambda}=-\frac{1}{3}-\frac{2\sqrt{\Omega_{\rm
\Lambda}}}{3c}\cos y+\frac{b^2(1+\Omega_{k}-\Omega_{\rm
m})}{\Omega_{\rm \Lambda}}.
\end{equation}
 From
Eqs.(\ref{eff}, \ref{stateq}), we have the effective equation of
state as
\begin{equation} \label{3eq401}
w_{\rm \Lambda}^{eff}=-\frac{1}{3}-\frac{2\sqrt{\Omega_{\rm
\Lambda}}}{3c}\cos y.
\end{equation}
If we take $c=1$, then $w_{\rm \Lambda}^{eff}$ is bounded from below
by \be w_{\rm \Lambda}^{eff}=-\frac{1}{3}(1+2\sqrt{\Omega_{\rm
\Lambda}}) \label{bond} \ee taking $\Omega_{\Lambda}=0.73$ for the
present time, the lower bound of $w_{\rm \Lambda}^{eff}$ is
$-0.9$.\footnote{The differential equation for $\Omega_{\Lambda}$ is
as
\begin{equation}
\frac{d\Omega_{\Lambda}}{dx}=2\Omega_{\Lambda}(\frac{\sqrt{\Omega_{\Lambda}}}{c}\cos
y+q)
\end{equation}
where $x=\ln a$, and
\begin{equation}
q=-1-\frac{\dot{H}}{H^2}
\end{equation}
Differentating Eq.(17) with respect to cosmic time $t$ and then
using Eqs.(16,18) one have
 \begin{equation}
\dot{H}=\frac{-4\pi G_{4}}{3}[3\rho_{\Lambda}(1+w_{\rm
\Lambda})+3\rho_{m}+4\chi]-\frac{k}{a^{2}}.
\end{equation}
 } Therefore it is impossible
to have $w_{\rm \Lambda}^{eff}$ crossing $-1$. This implies that one
can not generate phantom-like equation of state from an interacting
holographic dark energy model in non-flat universe. As it was
mentioned in introduction, $c$ is a positive constant in holographic
model of dark energy, and($c\geq1$). However, if $c<1$, the
holographic dark energy will behave like a Quintom model of DE
.\cite{quintom}, the amazing feature of which is that the equation
of state of dark energy component $w_{\rm \Lambda}$ crosses $-1$.
Hence, we see, the determining of the value of $c$ is a key point to
the feature of the holographic dark energy and the ultimate fate of
the universe as well. However, in the recent fit studies, different
groups gave different values to $c$. A direct fit of the present
available SNe Ia data with this holographic model indicates that the
best fit result is $c=0.21$ \cite{HG}. Recently, by calculating the
average equation of state of the dark energy and the angular scale
of the acoustic oscillation from the BOOMERANG and WMAP data on the
CMB to constrain the holographic dark energy model, the authors show
that the reasonable result is $c\sim 0.7$ \cite{cmb1}. In the other
hand, in the study of the constraints on the dark energy from the
holographic connection to the small $l$ CMB suppression, an opposite
result is derived, i.e. it implies the best fit result is $c=2.1$
\cite{cmb3}.
 \section{Conclusions}
 It is of interest to remark that in the literature, the different
scenarios of DE has never been studied via considering special
similar horizon, as in \cite{davies} the apparent horizon, $1/H$,
determines our universe. As we discussed in introduction for flat
universe the convenient horizon looks to be event horizon, while in
non flat universe we define $L$ because of the problems that arise
if we consider event horizon or particle horizon (these problems
arise if we consider them as the system's IR cut-off). Thus it looks
that we need to define a horizon that satisfies all of our accepted
principles; in \cite{odintsov} a linear combination of event and
apparent horizon, as IR cut-off has been considered. In present
paper, we studied $L$, as the horizon measured from the sphere of
the horizon as system's IR cut-off. Then by considering the effects
of the interaction between a brane universe and the bulk
 we have obtained the equation of state for
the interacting holographic energy density in the non-flat universe.
Our calculation show, taking $\Omega_{\Lambda}=0.73$ for the present
time, the lower bound of $w_{\rm \Lambda}^{eff}$ is $-0.9$.
Therefore it is impossible to have $w_{\rm \Lambda}^{eff}$ crossing
$-1$. This implies that one can not generate phantom-like equation
of state from an interacting holographic dark energy model in
non-flat universe.

\end{document}